\documentclass[english,12pt,sort&compress,jpa]{iopart}
\usepackage[T1]{fontenc}
\usepackage[latin1]{inputenc}
\usepackage{float}
\usepackage{iopams}
\usepackage{setstack}
\usepackage{graphicx}
\usepackage{amssymb}
\usepackage[numbers]{natbib}
\usepackage{verbatim}
\usepackage{color}


\usepackage{babel}

\makeatother

\newcommand{\eqs}{Eqs.\ }
\newcommand{\Eqs}{Eqations\ }
\newcommand{\figs}{Figs.\ }

\begin{document}

\title[]{Interface localization in the 2D Ising model with a driven line}

\author{O Cohen and D Mukamel}

\address{Department of Physics of Complex Systems, Weizmann Institute of Science, Rehovot 76100, Israel}

\ead{\mailto{or.cohen@weizmann.ac.il} and \mailto{david.mukamel@weizmann.ac.il}}

\begin{abstract}
We study the effect of a one-dimensional driving field on the interface between two coexisting phases in a two dimensional model.
This is done by considering an Ising model on a cylinder with Glauber dynamics in all sites and additional biased Kawasaki dynamics in the central ring.
Based on the exact solution of the two-dimensional Ising model, we are able to compute the phase diagram of the driven model within a special limit of fast drive and slow spin flips in the central ring. The model is found to exhibit two phases where the interface is pinned to the central ring: one in which it fluctuates symmetrically around the central ring and another where it fluctuates asymmetrically.
 In addition, we find a phase where the interface is centered in the bulk of the system, either below or above the central ring of the cylinder. In the latter case, the symmetry breaking is `stronger' than that found in equilibrium when considering a repulsive potential on the central ring. This equilibrium model is analyzed here by using a restricted solid-on-solid model.

\end{abstract}



\maketitle

\section{Introduction}


%

The properties of an interface separating two coexisting phases in two-dimensions has been studied extensively in the past.
Many of these studies deal with the wetting transition between a `dry' phase where the interface is attached to the boundary
and a `wet' phase where it is found in the bulk. This transition has been observed in various experimental settings (see review in \cite{sullivan1986wetting}),
as well as studied theoretically (see review in \cite{binder2003monte} and various studies in \cite{abraham1980solvable,fisher1981scaling,nakanishi1982multicriticality,nakanishi1983critical,binder1986critical,
abraham1986exactly,binder1988wetting,binder1989wetting,albano1989adsorption,albano1990critical,parry1990influence,
parry1992novel,binder1995character,binder1995thin,binder1996interface,maciol1996d,maciolek1996magnetization,carlon1998critical,pfister1999interface,
albano2000monte,de2005interfaces,schulz2005first,pang2011simulation}).
In most theoretical models considered, when the system is its wet phase, the interface performs a random walk in the bulk of the system in a potential that becomes flat in the thermodynamic limit. An interesting question, with significant experimental implications, is whether and how the interface can be confined to a certain region of the system or pinned to a specific line.

One of the prominent prototypical models studied in this context is the two-dimensional Ising model with Glauber (spin-flip) dynamics.
The phase diagram of the equilibrium Ising model with cylindrical boundary conditions has been analyzed in the past using its exact solution \cite{McCoy,abraham1980solvable}. In the ordered phase and for boundary magnetic fields that are of opposite signs, the model exhibits an interface
between a positive and a negative magnetic phases. When the weaker of the two boundary fields is of small enough amplitude, the
model is in a `dry' phase where the interface is attached to this boundary. As this boundary field increases, the model undergoes
a second order wetting transition where the interface detaches from the boundary and moves to the bulk of the system.
This transition has been studied analytically in more general geometries in \cite{pfister1999interface,abraham2000orientation}.

One natural way to confine the interface in the bulk is to modify the interaction strength of the model in a single line. This approach has been considered in  \cite{abraham1981binding,van1981pinning}, by studying
 an Ising model on a cylinder where the interaction strength within the central ring of the cylinder, denoted by $J_0$, is smaller than that in the bulk, denoted by $J$.
 As often done in the study of wetting phenomena, the full two-dimensional model has been approximated in these studies by a one-dimensional restricted solid on solid (RSOS) model, which is more analytically tractable.
 The RSOS model describes the position
of a sharp interface between pure positive and negative magnetic phases. The term `pure' refers here to the absence of droplets of opposite sign in each phase. This description thus captures the low temperature behaviour of the full two dimensional Ising model.
For $J_0<J$ and with boundary fields that are strong enough to induce a wet phase, it was found that the interface is pinned to the central ring and fluctuates symmetrically around it.
Here we complement this analysis by considering also the case where $J_0>J$. In this case we find a phase where the interface is repelled from the central ring and confined to one of its sides.
Note that here and below we use the terms `dry' and `wet'  to describe the effect of the boundary layers, as opposed to `pinned' and `repelled', which describe the effect of the central ring.

The aim of this paper is to study a different mechanism for confining an interface, obtained by exerting a local driving field,
which results in a non-equilibrium steady-state.
We study this mechanism by considering the example of the Ising model on a cylinder with Glauber dynamics (spin-flip)
in the bulk and additional biased Kawasaki dynamics (spin-exchange) in the central ring of the cylinder.

In general, the addition of the biased Kawasaki dynamics to the equilibrium model results in a non-equilibrium steady-state measure which is unknown.
However, as shown below, in the specific limit of fast drive and slow spin flips in the central ring of the cylinder, the phase diagram of the model can be studied analytically.
 Based on the approach presented in \cite{Cohen2012} and employed for the magnetization-conserving version of our model in \cite{Sadhu2012},
 the dynamics of the magnetization in the central ring, $m_0$, can be shown to correspond in this limit to an equilibrium random walk in a potential.
This simplification is due to the fact that between every spin flip in the central ring, the bulk of the system above and below this ring relaxes to the steady-state
of the two-dimensional Ising model. The effective potential, within which $m_0$ evolves, can be derived from the exact solution of the 2D Ising model, presented in \cite{McCoy}. For the purpose of our analysis this solution, originally presented for the case of a single boundary field,
 is generalized below to the case of two boundary fields.



The steady-state distribution of the random walker yields the large deviation function of $m_0$, from which the phase diagram of the model can be derived.
In addition to the pinned phase and repelled phase observed in the equilibrium RSOS model,
we find an {\it asymmetrically-pinned phase} where the sign-flip symmetry of $m_0$ is spontaneously broken. The interface is pinned in this case to the central ring but fluctuates asymmetrically around it.

One important difference between the equilibrium and non-equilibrium cases involves the nature of the repelled phase.
In both cases, the repelled phase is composed of two equally probable states where the interface is either below or above the central ring. In equilibrium, the system switches
between these two states on a time-scale which grows polynomially with the height and circumference of the cylinder. By contrast, in the driven case, this time scale is found to grow exponentially with the circumference of the cylinder. We thus refer to the latter phase as a `strongly repelled phase'. This strong repulsion is assumed to be due to long-range
correlations cause by the driving field, which are known to exist generically in driven systems \cite{Spohn1983,Garrido1990,Dorfman1994,domb1995statistical,Schmittmann1995,lepri2003thermal,Ortiz2004,derrida2007non,blythe2007nonequilibrium,dhar2008heat,Sadhu2011}.

The asymmetrically-pinned phase has also been observed by Sadhu et al. who studied the same model considered here, only with {\it conserving} Kawasaky dynamics in the bulk \cite{Sadhu2012}.
The original motivation for this work are experimental results on colloidal gas-liquid interface subjected to a shear flow
parallel to the interface \cite{derks2006suppression}. These experiments have shown that the shear drive applied
away from the interface strongly suppresses the fluctuations of the interface, making it smoother.
As found below, the driving field, considered in \cite{Sadhu2012}, causes
a spontaneous breaking of the sign-flip symmetry of $m_0$, yielding a phase where the interface exhibits stronger fluctuations in one of the sides of the central ring.
The analytic analysis of this {\it conserving} model was possible for strong drive, slow spin flips on the central ring and low temperature.
 In the present study, we analyze the {\it non-conserving} model in a similar limit but for finite temperatures.
  We are thus able to gain a better understanding of the fluctuations of the interface observed numerically in \cite{Sadhu2012}.

We complement the theoretical study by numerical simulation of the Ising model with a driven line. Interestingly, the two pinned phases and the strongly-repelled phase, found analytically for fast drive and slow spin-flips on the central ring, are found to exist significantly away from this limit.


The paper is organized as follows: In \sref{sec:model} we define the Ising model with a driven line.
In \sref{sec:ising} we discuss some of the properties of the equilibrium Ising model with $J_0=J$, obtained from the exact solution of the 2D Ising model \cite{McCoy}. These properties are used in the derivation of the non-equilibrium model and provide a context for our present work.
The properties of the equilibrium model for $J_0\neq J$ are analyzed using a solid-on-solid model in \sref{sec:interface}.
The analysis of the non-equilibrium Ising model with a driven line is carried out in \sref{sec:driven_ising}.
\Sref{sec:numerical} presents results of Monte Carlo simulations which support some of the results of \sref{sec:driven_ising}. Concluding remarks are provided in \sref{sec:conc}.





\section{The Ising model on a cylinder with a driven line}
\label{sec:model}
\begin{figure}
\noindent
\begin{centering}\includegraphics[scale=2.0]{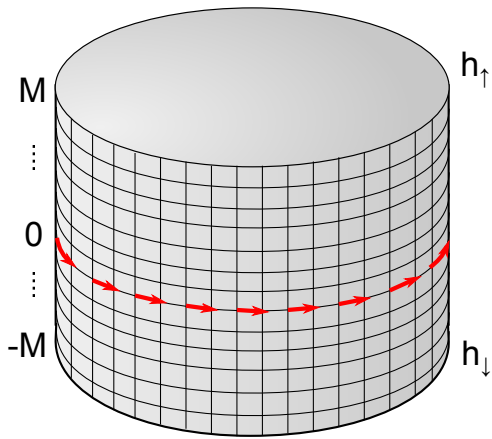}\par\end{centering}
\caption{
\label{fig:ising}
 Cylindrical lattice with the drive on the central ring and the boundary fields indicated.}
\end{figure}

The Ising-type model studied below is defined on a cylinder of size $L\times (2M+1)$, illustrated in \fref{fig:ising}. A micro-configuration of the model is denoted by
 $\boldsymbol \sigma \equiv \{ \sigma_{i,j} \}$, where $i\in[1,L]$, $j\in[-M,M]$ and $\sigma_{i,j}=\pm 1$.
Every spin in the system evolves by Glauber dynamics defined for spin $i,j$ as
\begin{equation}
\label{eq:ising_bulk}
\boldsymbol \sigma \qquad \overset{p_{i,j}^{(\mathrm{Gb})} e^{-\frac{1}{2}\beta \{ H[\boldsymbol \sigma^{(i,j)} ]-H[\boldsymbol \sigma]\} }}{\longrightarrow} \qquad \boldsymbol \sigma^{(i,j)},
\end{equation}
where $\boldsymbol \sigma^{(i,j)}$ is the original spin-configuration obtained after flipping the spin $(i,j)$ in $\boldsymbol \sigma$.
Here $p_{i,j}^{(\mathrm{Gb})}$ denotes the local rate of the Glauber dynamics.
These rates affect the steady-state of the model only when a driving field is present. In our analysis of the non-equilibrium steady-state of the model in \sref{sec:non_eq}
we consider the limit where the dynamics on the central ring is slower, i.e. $p_{i,0}^{(\mathrm{Gb})}\ll p_{i,j\neq 0}^{(\mathrm{Gb})}=1$.
 The Hamiltonian in \eref{eq:ising_bulk} is defined as
\begin{eqnarray}
\label{eq:ising_H}
 H[\boldsymbol \sigma]&=&-J\sum_{i=1}^L \sum_{j=-M}^{j=M-1} ( \sigma_{i,j}\sigma_{i,j+1} + \sigma_{i,j}\sigma_{i+1,j} ) \\
&&-(J_0-J)\sum_{i=1}^L  \sigma_{i,0}\sigma_{i+1,0}
-\sum_{i=1}^L ( h_{\downarrow} \sigma_{i,-M} + h_{\uparrow} \sigma_{i,M}), \nonumber
\end{eqnarray}
where $J$ denotes the coupling strength in the bulk, $J_0$ is the coupling strength within the central ring
and $h_{\downarrow},h_{\uparrow}$ are the boundary fields at the bottom and top of the cylinder, respectively (see \fref{fig:ising}).
The periodic boundary condition is taken into account in \eref{eq:ising_H} by considering $\sigma_{L+1,j}=\sigma_{1,j}$.

In addition to the Glauber dynamics in \eref{eq:ising_bulk}, the model is driven out of equilibrium by Kawasaki dynamics on the central ring defined by the following rates:
\begin{equation}
\label{eq:ising_drive}
+- \overset{p^{(\mathrm{Kw})}q} {\underset{p^{(\mathrm{Kw})} (1-q)}{\leftrightarrows}} -+,
\end{equation}
where $q$ controls the asymmetry of the dynamics and $p^{(\mathrm{Kw})}$ is the rate of the Kawasaki dynamics.
 It is important to note that even for $q=1/2$ the model is out of equilibrium.
The equilibrium Ising model is retained only when considering $p^{(\mathrm{Kw})}=0$.

In section \sref{sec:eq} we discuss the properties of the equilibrium model ($p^{(\mathrm{Kw})}=0$). In the case where $J_0=J$ the model reduces to the standard 2D Ising model which has been studied analytically by many authors. When the boundary fields are of opposite sign the model exhibits an interface between a plus phase and a minus phase.
The model undergoes a second order wetting transition when $T$ is fixed and the boundary fields are increased ( or conversely when the boundary fields are fixed and $T$ is decreased). For $J_0\neq J$, we study the model
by considering the corresponding restricted solid on solid (RSOS) model, which is found to exhibit a phase where the interface is pinned to the central ring for $J_0<J$ and a phase where it is `weakly repelled' from the central ring for $J_0>J$. The terms weakly is clarified below.
The non-equilibrium case is analyzed analytically in \sref{sec:ising} in the limit of fast driven and slow spin flips in the central ring, $p_{i,0}^{(\mathrm{Gb})}\ll p_{i,j\neq 0}^{(\mathrm{Gb})} \ll p^{(\mathrm{Kw})}$. The qualitative picture, which emerges from the analytic analysis, is confirmed numerically in \sref{sec:numerical}.




\section{Analysis of the equilibrium Ising model}
\label{sec:eq}

\subsection{Wetting transition in the equilibrium model with $J_0=J$}
\label{sec:ising}

In this section we review the phase diagram of the equilibrium Ising model ($J_0=J$ and $p^{(\mathrm{Kw})}=0$), as obtained using the exact solution of the model, presented in the treatise by McCoy and Wu \cite{McCoy}. The authors have studied the Ising model on a cylinder with a single boundary field, e.g. for $h_\uparrow=0$. In \ref{sec:exact} we generalize their solution to the case of two boundary fields, where one finds a non-trivial phase diagram.
  This phase diagram can also be obtained based on the analysis of Abraham in \cite{abraham1980solvable}. We chose, however, to use the solution in \cite{McCoy} since it readily yields an expression for the boundary magnetization (of the top and bottom rings), necessary for the analysis in \sref{sec:driven_ising}.

The partition function of the equilibrium Ising model on a cylinder is given by
\begin{equation}
Z=\sum_{\boldsymbol \sigma} e^{-\beta H[\boldsymbol \sigma]},
\end{equation}
where $ H[\boldsymbol \sigma]$ is defined in \eref{eq:ising_H} and $J_0$ is taken in this section to be $J_0=J$.
In \cite{McCoy} the logarithm of this function is computed explicitly for $h_\uparrow=0$ and written as
\begin{equation}
\label{eq:ising_free_energy}
-\frac{1}{\beta}\ln Z= 4 L (2M+1) F + 2 L \mathcal{F}_0 +  L \mathcal{F}(h_\downarrow)+\mathcal{O}(1).
\end{equation}
Here $F$ is the well-known bulk-free-energy of the Ising model, $\mathcal{F}_0$ is the free energy of each boundary in the absence of a magnetic field
and $\mathcal{F}(h_\downarrow)$ is the leading order contribution in $L$ of the magnetic field to the free energy. These free energies are defined in terms
of integrals over an angle $\theta$, which result from the Fourier transform over the horizontal coordinate. More details are provided in section VI of \cite{McCoy}
and in \ref{sec:exact}. The bulk-free-energy is given by
\begin{equation}
\label{eq:Fbulk_def}
F=-\frac{1}{\beta} \Big \{ \ln [ 2 \cosh^2 ( \beta J)]+\frac{1}{4\pi} \int_{-\pi}^{\pi} d\theta \ln[  z_J(1-z_J^2) \alpha] \Big \}
\end{equation}
where $z_J\equiv \tanh(\beta J)$ and $\alpha$ is the solution with the larger magnitude of the following quadratic equation:
\begin{equation}
\label{eq:alpha_def}
(1+z^2_J)^2-2z_J(1-z^2_J)\cos\theta -z_J(1-z^2_J)(\alpha+\alpha^{-1})=0.
\end{equation}
The boundary free energy in the absence of a boundary field is given in terms of $\alpha$ by
\begin{equation}
\fl \mathcal{F}_0 =-\frac{1}{\beta} \Big \{ -\ln [ \cosh(\beta J)] -\frac{\ln 2}{2} +\frac{1}{4\pi} \int_{-\pi}^{\pi} d\theta \ln \big[ 1+ \frac{1-z_J^4-2z_J(1+z_J^2)\cos \theta}{z_J(1-z_J^2)(\alpha-\alpha^{-1})} \big ].
\end{equation}
The field-dependent boundary free energy, which is studied extensively below, is given by
\begin{equation}
\label{eq:ising_fh}
\fl \mathcal{F}(h)= -\frac{1}{\beta} \Big\{ \ln [ \cosh(\beta h)] +\frac{1}{4\pi } \int_{-\pi}^\pi d\theta \ln \big[
1- \frac{z_h |1+e^{i\theta}|^2}{z_J(1+z^2_J+2z_J\cos \theta)-(1-z^2_J)\alpha} \big]\Big\}
\end{equation}
where $z_h\equiv \tanh(\beta h)$.

In \ref{sec:exact} we show that the free energy for $h_\uparrow\neq0$ and $h_\downarrow\neq 0$ is given by
\begin{equation}
\label{eq:ising_F1}
-\frac{1}{\beta}\ln Z= 4 L (2M+1) F + 2 L \mathcal{F}_0 +  L \mathcal{F}(h_\downarrow)+  L \mathcal{F}(h_\uparrow)+\mathcal{O}(1).
\end{equation}
The fact that the $\mathcal{O}(L)$ term is simply a sum over those in \eref{eq:ising_free_energy} over the two fields, $h_\downarrow$ and $h_\uparrow$,
suggests that the two boundaries affect the system independently. However, there are coupling terms between the fields in the $\mathcal{O}(1)$ terms, which
determine the overall magnetization of the system, i.e. the position of the interface.

\begin{figure}
\noindent
\begin{centering}\includegraphics[scale=0.6]{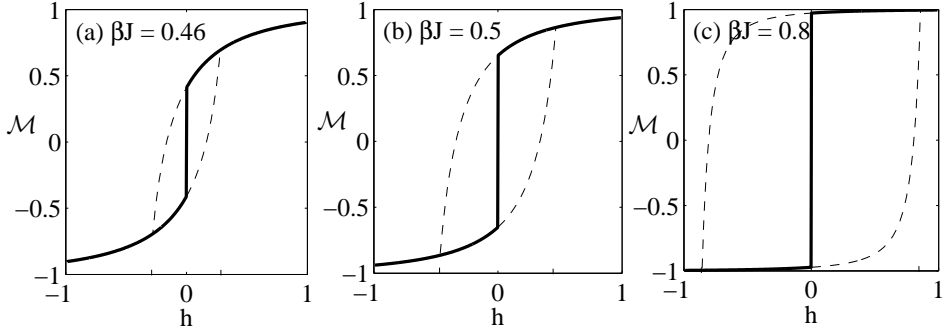}\par\end{centering}
\caption{
\label{fig:mbh} The boundary magnetization in the two-dimensional model as a function of the magnetic field. The thick solid and dashed line
denote $\mathcal{M}(h)$ and $\mathcal{M}_{d}(h)$, respectively.
}
\end{figure}

In the analysis of the of the driven Ising model, defined in \sref{sec:model}, we use the expression for the
boundary magnetization of the equilibrium model, obtained from the derivative of $\mathcal{F}(h)$ as
\begin{eqnarray}
\label{eq:ising_mh}
\fl \quad \mathcal{M}(h)&=&-\frac{d \mathcal{F}(h)}{dh} \\
\fl &=& z_h+z_h \frac{z(1-z_J^2)}{2\pi} \int_{-\pi}^\pi d\theta \frac {|1+e^{i\theta}|^2}{z_h^2 z_J |1+e^{i\theta}|^2 -z_J^2(1+z_J^2+2z_J\cos\theta)+z_J(1-z_J^2)\alpha} \nonumber
\end{eqnarray}
The properties of $\mathcal{M}(h)$ are analyzed in detail in Sec. VI.5 of \cite{McCoy} for the case of a single boundary field, obtained here by setting $h_\uparrow=0$. In this case,
below the Ising critical temperature, $T_c\approx 2.269 J/k_B$, the boundary magnetization, $\mathcal{M}(h_\downarrow)$, exhibits a discontinuity at $h_\downarrow=0$ which signifies a sign flip of the bulk magnetization. This discontinuity is evident in \fref{fig:mbh},
where $\mathcal{M}(h)$ is denoted by the thick solid line. McCoy and Wu have computed an additional solution for the boundary-magnetization using an analytic continuation of the integral in \eref{eq:ising_mh}, which is given by
\begin{equation}
\mathcal{M}_{d} (h)=\mathcal{M}(h) + \frac{2z_h [(1+z_J)^2(z^2_J-z_h^2)- z_J^2 (1-z_J)^2(1-z_h^2)^2]}{(r^{-1}-r)(1-z_h^2)z_J(z_J^2-z_h^2)^2}.
\end{equation}
Here $r$ denotes either of the two solutions of the following quadratic equation:
\begin{equation}
r+r^{-1}=\frac{2 z_J^3 -2 z^4_h z_J - z^2_J (1 - z_J^4)}{z_J(1 - z^2_h)(z_J^2-z_h^2)}.
\end{equation}
This analytic continuation of $\mathcal{M}(h)$, denoted by the two dashed lines in \fref{fig:mbh}, is valid only in values of $h$ where $r$ is real, namely for $h\in [-h_w(T),h_w(T)]$, where
\begin{equation}
\label{eq:htilde}
h_w(T)=\frac{1}{\beta} \, \mathrm{arctanh} \Big[ \sqrt{ 2+z_J-2\frac{1+z_J}{1+z_J^2}} \, \Big].
\end{equation}
The subscript $d$ in $\mathcal{M}_{d}$ signifies that  $\mathcal{M}_{d}$ is the magnetization when the system is in the dry phase and the interface is attached to the boundary. This interpretation is discussed below. Similarly $h_w(T)$ is the field where the wetting transition occurs.

 In \cite{McCoy}, where the case of $h_\uparrow=0$ was studied, $\mathcal{M}_{d} (h_\downarrow)$ is interpreted as a
 metastable solution which forms a hysteresis loop of the boundary magnetization. The  hysteresis loop is visible
 in \fref{fig:mbh} when combining the dashed and solid lines. When the system is on the dashed line the boundary field, $h_\downarrow$, and bulk magnetization are of opposite signs. In this case
 there exists a phase with a lower free energy, where the bulk magnetization switches sign and the interface moves to the other boundary where $h_\uparrow=0$. The transition between those two states involves creating an
 interface in the bulk of the system. The free energy barrier between the metastable state and the ground state can thus be interpreted as the free energy difference between the case where the interface is attached to
  a boundary with a field $h_\downarrow$ and the case where it is in the bulk. This barrier vanishes for $|h_\downarrow|=|h_w(T)|$, where the dashed and solid lines in \fref{fig:mbh} meet.

 In the case of where there are two non-vanishing boundary fields, \eref{eq:ising_F1} implies that the magnetization in the top and bottom lines should still be given by either $\mathcal{M}(h)$ or $\mathcal{M}_d(h)$ with $h=h_\uparrow,h_\downarrow$, respectively. However, the interpretation of each branch should be different in this case.
 When the two fields are of opposite sings, the state where the interface is attached to the field with the lower magnitude has
the lower free energy out of the two dry states. We consider for concreteness the case where indeed $h_\uparrow  h_\downarrow<0$ and $|h_\downarrow|<|h_\uparrow|$.
Using the above interpretation of the free energy barrier of the metastable state, it is clear that for $|h_\downarrow|<h_w(T)$
 the state where the interface is attached to the lower boundary has a lower free energy than that of the wet state. In this case the magnetization of the
 lower boundary is given by $\mathcal{M}_d(h_\downarrow)$. On the other hand when $|h_\downarrow|>h_w(T)$ the state with the lowest free energy is that where the interface is found in the bulk. The boundary magnetization is given in this case by $\mathcal{M}(h_\downarrow)$.  One can therefore  write the magnetization of the lower boundary as
  \begin{equation}
\label{eq:muhh}
\fl \qquad \mathcal{M}(h_{\downarrow},h_{\uparrow})=\left\{ \begin{array}{ccc}
\mathcal{M}_d(h_{\downarrow}) & \qquad & |h_{\downarrow}|<\min(h_w(T),|h_\uparrow|) \quad \mathrm{and} \quad h_{\downarrow}h_{\uparrow}<0\\
\mathcal{M}(h_{\downarrow}) &  \qquad & \mathrm{else}
\end{array}\right. ,
\end{equation}
Similarly, the expression for the upper-boundary magnetization can be obtained by exchanging $h_\uparrow$ and $h_\downarrow$ in the above equation.

\begin{figure}
\noindent
\begin{centering}\includegraphics[scale=1]{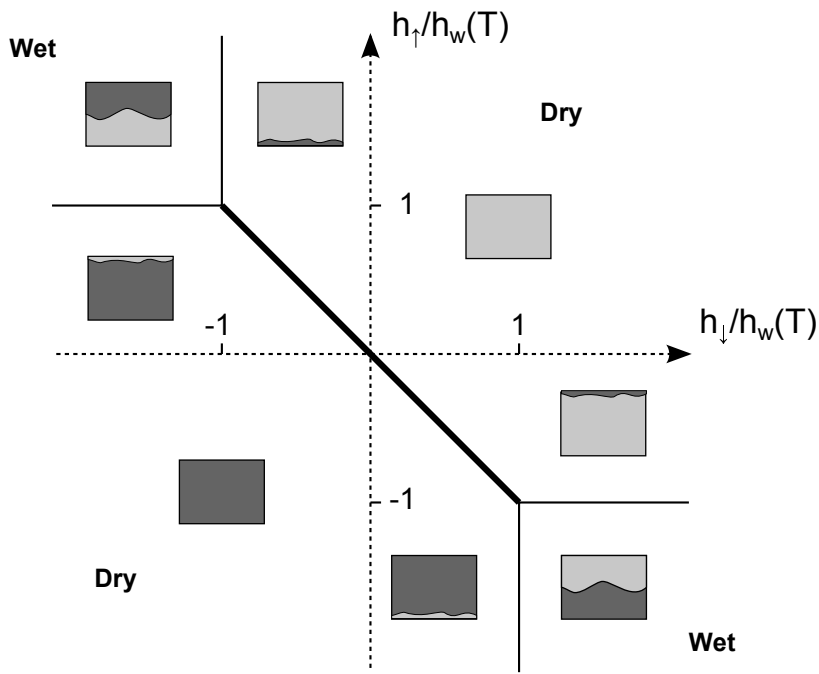}\par\end{centering}
\caption{
\label{fig:pinning_diagram}
Phase diagram of the Ising model on a cylinder with two boundary fields, $h_\downarrow$ and $h_\uparrow$, for a fixed temperature below $T_c$.
The thick and thin solid lines denote first and second order transition lines, respectively.
 The state of the system is sketched in boxes that signify a typical spin state in the cylinder, where
light gray denotes the plus phase and dark gray denotes the minus phase. }
\end{figure}

Based on \eref{eq:muhh} we can draw the phase diagram of the Ising model on a cylinder with two boundary fields, shown for a fixed temperature below $T_c$ in \fref{fig:pinning_diagram}.
The thin solid line denote a second order wetting transition, between a state where the interfaces is attached to one of the boundaries and a state
where it is found in the bulk of the system. The thick solid line denotes a first order transition where the interface shifts from one boundary to the other.
It is important to note that the bulk magnetization changes smoothly across the second order transition line. Close to the transition line the interface is not in the middle of the cylinder, but
closer to one of the boundaries (although still $O(L)$ sites away from each). The exact position of the boundary can in principle be computed from the $\mathcal{O}(1)$ term in \eref{eq:ising_F1} which couples $h_\downarrow$ and $h_\uparrow$ explicitly.

One should note that the above interpretation of $\mathcal{M}_d(h)$ for the case of two boundary fields, from which \eref{eq:muhh} and the phase diagram given in \fref{fig:ising_phase} are derived, is done here on a heuristic level. A similar
  connection between $\mathcal{M}_d(h)$ and the wetting transition has already been mentioned in \cite{pfister1987phase,pfister1999interface}.
  A full mathematical proof  of \eref{eq:muhh} based on the Pfaffian method employed by McCoy and Wu is beyond the scope of this paper.

\begin{figure}
\noindent
\begin{centering}\includegraphics[scale=0.6]{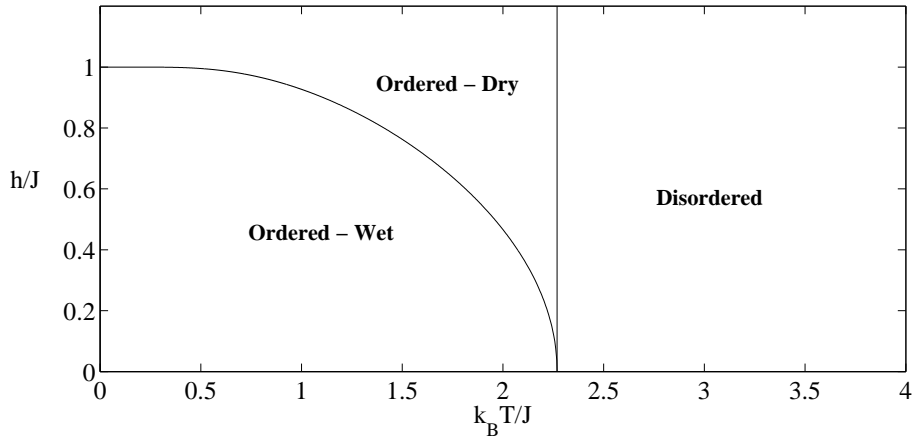}\par\end{centering}
\caption{
\label{fig:singlefield} Two dimensional projection of the phase diagram of the Ising model on a cylinder onto the $h_\uparrow=-h_\downarrow$ plane.
 The solid lines denote second order transitions between the
three phases marked by the labels. The ordering of the bulk system occurs below the known Ising critical temperature, $T_c \approx 2.269J/k_B$.
 The wetting transition line occurs at $|h_\uparrow|=|h_\downarrow|=h_w(T)$ where $h_w(T)$ is given in \eref{eq:htilde}.
}
\end{figure}

It is also instructive to project the phase diagram of the model from the $h_\uparrow,h_\downarrow,T$-space onto the $h_\uparrow=-h_\downarrow$ plane,
as shown in \fref{fig:singlefield}. The diagram shows the ordered-disordered second order transition at the bulk critical temperature and a wetting transition
at $h_\downarrow=h_\uparrow=h_w(T)$. The wetting temperature, $T_w$, has been derived in the past in \cite{abraham1980solvable}.
This transition has been studied analytically also for different geometries in \cite{pfister1999interface,abraham2000orientation}.

\subsection{RSOS analysis for $J_0\neq J$}
\label{sec:interface}
In general, the solution of the equilibrium model with $J_0\neq J$ is unknown. However, as often done in the analysis of wetting phenomena, the qualitative behaviour of model can be analyzed by considering the corresponding
restricted solid on solid (RSOS) model. Because RSOS models assume the two bulk phases to be pure (without droplets of phases of the opposite sign), they can be thought of as describing the low temperature behaviour of their corresponding
two dimensional versions. In our case, the interface is described by an interface height variable, $h_i\in [-M,M+1]$, such that all the spin above (below) $h_i$ are assumed to be of sign equal to that of $h_\uparrow$ ($h_\downarrow$). The Hamiltonian of the RSOS model corresponding to the Hamiltonian in \eref{eq:ising_bulk} is given by
\begin{equation}
\label{eq:ham_interface}
H=LJ+J\sum_{i=1}^{L}\delta_{|h_{i+1}-h_{i}|-1}+(J_{0}-J)[\delta_{h_{i}}\delta_{h_{i+1}-1}+\delta_{h_{i}-1}\delta_{h_{i+1}}],
\end{equation}
and in addition $H=\infty$ whenever $|h_{i+1}-h_{i}|>1$, which maintains the continuity of the interface. In this paper we are interested mainly in the dynamics of the central ring affects the wet phase. It is therefore sufficient to consider here the case where $h_\downarrow \to -\infty, h_\uparrow \to \infty$, in which the interface cannot be attached to the wall. The effect of the boundary fields can in principle be analyzed using the method described below.

Since the RSOS model is one-dimensional, its analysis can be done using the transfer matrix method. The transfer matrix corresponding to \eref{eq:ham_interface} is of dimension $(2M+2)\times(2M+2)$
and is given by
\begin{equation}
T_{i,j}=\left\{ \begin{array}{cc}
e^{-\beta[J+J\delta_{|i-j|-1}+(J_{0}-J)(\delta_{i}\delta_{j-1}+\delta_{i-1}\delta_{j})]} & \qquad|i-j|\leq1\\
0 & \mathrm{else}
\end{array}\right. ,
\end{equation}
where $i,j\in[-M,M+1]$. The partition function of the model can be computed using the eigenvalues of $T$, denoted by $\lambda_m$, as
\begin{equation}
Z=\mathrm{tr}[T^{L}] = \sum_{m=1}^{2M+2} \lambda^L_m.
\end{equation}
In addition, we denote by $|\psi^{(m)}\rangle$ the eigenvector corresponding to $\lambda_m$ and by $\psi_i^{(m)}$ its $2M+2$ components.

The properties of the different phases of the model can be understood by analyzing the probability distribution function (PDF) of $h_i$, which is of course independent of $i$. In order to compute $Pr(h_i=j)$, it is useful to define a basis of vectors, $|\phi^{(j)}\rangle$ with $j=-M, \dots M+1$, for which the $j$'th entry is $1$ and the rest are zero, i.e.
\begin{equation}
\label{eq:base_vector}
\phi_{\ell}^{(j)}\equiv \left\{ \begin{array}{ccc}
1 & \quad & \ell=j\\
0 & \quad & \mathrm{else}
\end{array}\right. .
\end{equation}
In terms of $|\phi^{(j)}\rangle$ the PDF of $h$ is given to using the density matrix approach as
\begin{equation}
\label{eq:Ph}
\fl Pr(h_{i}=j)=\frac{1}{Z}\langle \phi^{(j)}|T^{L}|\phi^{(j)}\rangle = \frac{1}{\sum _m\lambda^{L}}\langle\phi^{(j)}|\psi^{(m)}\rangle\lambda^{L}_m\langle\psi^{(m)}|\phi^{(j)}\rangle=\frac{\sum_m \lambda^{L}_m ( \psi_{j}^{(m)})^2}{\sum _m\lambda^{L}}.
\end{equation}
The PDF of $h$ is thus fully described by the eigenvectors and eigenvalues of $T$.
Analyzing the equation $T|\psi\rangle = \lambda |\psi\rangle$ for $\psi_j$ with $j\neq 0,1$, one find an equation of the form
\begin{equation}
\label{eq:psi_bulk}
\psi_{j-1}^{(m)}-2\psi_{j}^{(m)}+\psi_{j+1}^{(m)}=(\lambda_m e^{2\beta J}-e^{\beta J}-2)\psi_{j}^{(m)}.
\end{equation}
On the central lines the finite Laplacian operator in the LHS above is slightly modified, yielding
\begin{eqnarray}
\label{eq:psi_i1}
e^{-\beta(J_{0}-J)}\psi_{0}^{(m)}-2\psi_{1}^{(m)}+\psi_{2}^{(m)}&=&(\lambda_m e^{2\beta J}-e^{\beta J}-2)\psi_{1},\\
\label{eq:psi_i0}
\psi_{-1}^{(m)}-2\psi_{0}^{(m)}+e^{-\beta(J_{0}-J)}\psi_{1}^{(m)}&=&(\lambda_m e^{2\beta J}-e^{\beta J}-2)\psi_{0}.
\end{eqnarray}

\begin{figure}
\noindent
\begin{centering}\includegraphics[scale=0.7]{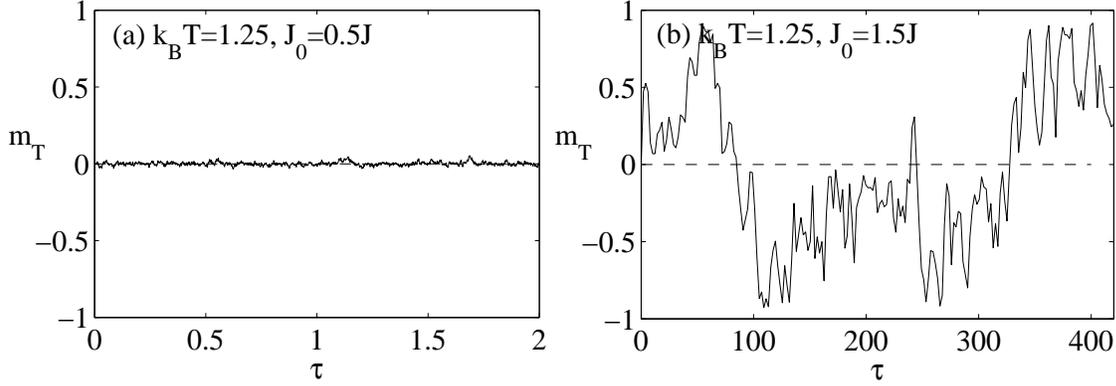}\par\end{centering}
\caption{\label{fig:mT_eq}
The overall magnetization in the system, $m_T(\boldsymbol \sigma) \equiv  \frac{1}{L(2M+1)} \sum_{i=1}^L\sum_{j=-M}^M\sigma_{i,j}$, as a function of a rescaled time variable
$\tau=t/[L(2M+1)]$ for $L=M=100$. The temperature in both figures is $k_B T=1.25$, which is below the Ising critical temperature. Figures (a) and (b) show the case where $J_0=0.5J$ and $J_0=1.5J_0$, respectively.
In (a) the interface is pinned symmetrically to the central ring, whereas in (b) the interface is weakly-repelled from the central ring.}
\end{figure}

Equations (\ref{eq:psi_bulk})-(\ref{eq:psi_i0}) can be solved for $J_0<J$ using the following Ansatz:
\begin{equation}
\psi_{i}^{(m)}=\left\{ \begin{array}{ccc}
A_m\sinh[k_m(M+1-i)] & \quad & 1\leq i \leq M+1\\
A_m\sinh[k_m(M+i)] & \quad & -M\leq i \leq 0
\end{array}\right.,
\end{equation}
where $k_m$ is a real exponent and $A_m$ is a normalization factor. Inserting this form into \eref{eq:psi_bulk} yields an expression for the eigenvalue, $\lambda_m = e^{-\beta J}+ 2e^{-2\beta J}\cosh k_m$.
Inserting this result and the ansatz into either \eref{eq:psi_i1} or \eref{eq:psi_i0} yields for the largest eigenvalue
\begin{equation}
k_1=\mathrm{acosh}[\frac{1+e^{2\beta(J-J_{0})}}{2e^{\beta(J-J_{0})}}]+O(e^{-M}),
\end{equation}
and additional solutions that are smaller than $k_1$ by an amount that does not scale with $M$. This gap between $\lambda_1$ and the second largest $\lambda$ implies that  the  properties of the system are thus fully controlled by the highest
eigenvalue of $T$.
The fact that $k_1$ does not scale with $M$ implies that the density profile of $h$ decays in the bulk of the system as
\begin{equation}
\frac{P(h_i=\alpha M )}{P(h_i=0)} \approx \Big( \frac{\psi_{\alpha M}^{(1)}}{\psi_0^{(1)}} \Big)^2 \sim e^{-Mk_1\alpha}.
\end{equation}
The interface is thus pinned to the central ring for $J_0<J$, with an exponentially small probability to be found in the bulk of the system.
The interface fluctuates symmetrically around the central ring because it is energetically preferable for the interface to cross it. We verify this numerically in the full two-dimensional model and for a specific temperature in \fref{fig:mT_eq}a.
 This phase is different from the asymmetrically pinned phase observed in the driven case below and in \cite{Sadhu2012}, where
the interface fluctuates more into one side of the central ring.

\begin{figure}
\noindent
\begin{centering}\includegraphics[scale=0.7]{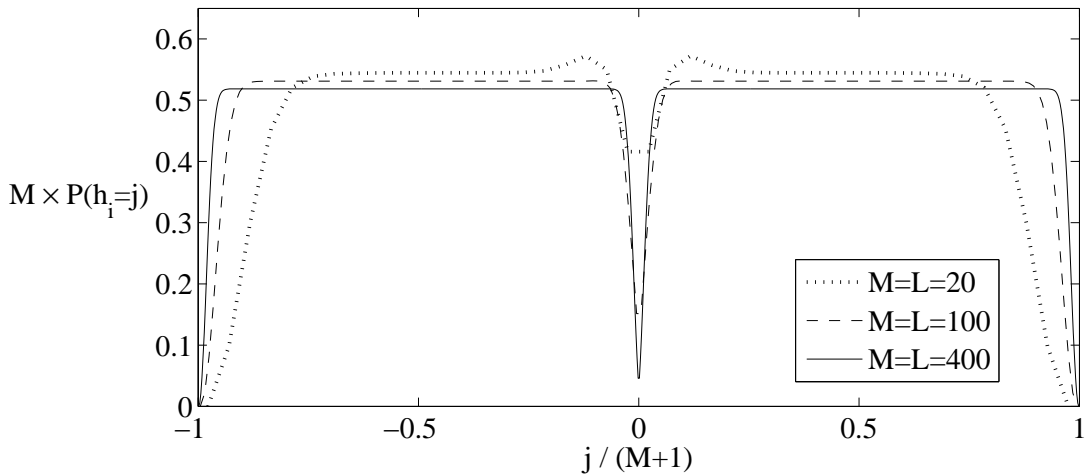}\par\end{centering}
\caption{\label{fig:interface_profile}
Numerical summation of the solution of $P(h_i=j)$ in \eqs (\ref{eq:Ph}),(\ref{eq:psi2}) and (\ref{eq:int_h20}) for $k_BT = 1.25$, $J_0=1.5J$ and $L=M=20,100,400$.
The probability rescaled by multiplication by $M$.
 }
\end{figure}

When analyzing the RSOS model with $J_0>J$, one has to consider a different solution for the eigenvectors of $T$ of the form
\begin{equation}
\label{eq:psi2}
\psi_{i}^{(m)}=\left\{ \begin{array}{ccc}
A_m\sin[k_m(M+1-i)] & \quad & 1\leq i \leq M+1 \\
A_m\sin[k_m(M+i)] & \quad &   -M\leq i \leq 0
\end{array}\right. ,
\end{equation}
where $k_m$ is a real wave-number.
Inserting the above ansatz into \eref{eq:psi_bulk} yields $\lambda_m = e^{-\beta J}+ 2e^{-2\beta J}\cos k_m$. Inserting the latter result and \eref{eq:psi2} into
either \eref{eq:psi_i1} or \eref{eq:psi_i0} yields
\begin{equation}
\label{eq:int_h20}
\cos(k_m)+\cot(k_m M)\sin(k_m)=e^{-\beta(J_{0}-J)}.
\end{equation}
In general one expects $k_m$ to scale as $1/M$ in order to yield a smooth density profile. As a result, there is no finite gap between the largest eigenvalue and others, which implies that many of the wavevectors ($O(M/\sqrt{L})$ of them) contribute to the density profile. Expanding the above equation in small $k_m$ yields $\cot(k_m M)k_m \approx e^{-\beta(J_{0}-J)}-1$,
which suggests that $k_m$ has the form $k=\frac{\pi m}{M}+\frac{\eta}{M^2}+O(M^{-3})$ with $m\neq0$. Inserting this expansion yields for $m\ll M$
\begin{equation}
\label{eq:int_h2}
k_m= \frac{\pi m }{M}-\frac{\pi m}{M^{2}(1-e^{-\beta(J_{0}-J)})}+O(M^{-3}),
\end{equation}
and thus $\eta=-\pi m / (1-e^{-\beta(J_{0}-J)})$.
Solution with $m\sim M$ may have a different form, but they do not contribute to the density profile due to the $\lambda_m^{L}$ factor in \eref{eq:Ph}.
Using this result the probability of the profile to touch the central ring can be shown to scale as
\begin{equation}
\label{eq:Ph0}
\fl \qquad \frac{P(h_i=0) }{P(h_i= \alpha M)} \sim  \frac{\sum_m \lambda_m^L \sin^2(k_m M)}{\sum_m  \lambda_m^L \sin^2[k_m (1-\alpha) M]} \approx
\frac{\sum_m \lambda_m^L \sin^2(\eta/M)}{\sum_m  \lambda_m^L \sin^2[(1-\alpha) \pi]} \sim \frac{1}{M^2},
\end{equation}
for any finite $\alpha$. The interface is thus repelled from the central. This is clearly visible in \fref{fig:interface_profile} where $P(h_i=j)$ is plotted
for a specific temperature and several system-sizes.

\begin{figure}
\noindent
\begin{centering}\includegraphics[scale=0.7]{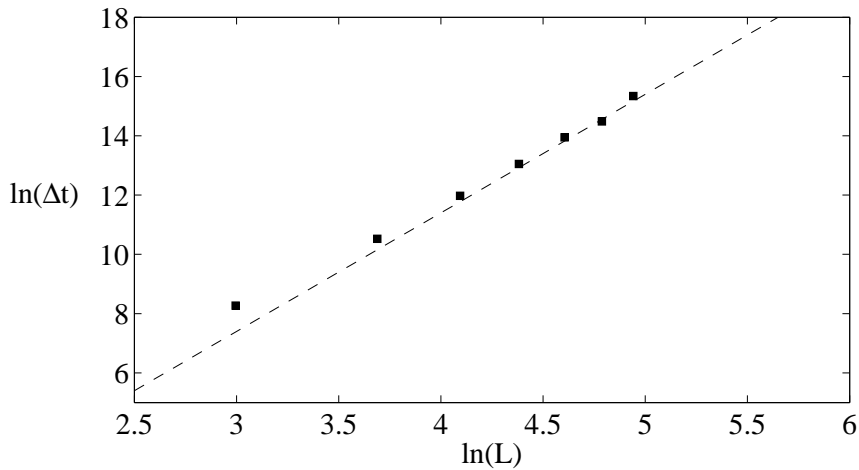}\par\end{centering}
\caption{\label{fig:switch_time_eq}
Average crossing time of the interface between the two sides of the central ring, plotted as a function of the length, $L$, on a log-log scale for $M=L$ and $k_BT=1.25$. The dashes line denotes
the curve $\Delta t \sim L^{4}$, which appears to fit the result.
 }
\end{figure}

In terms of temporal behaviour of the interface,  \eref{eq:Ph0} implies that as long as $L$ and $M$ scale similarly the interface is confined either above or below the central line and very rarely touches the central ring.  The is shown to be the case for the full 2D Ising model in \fref{fig:mT_eq}b, where $m_T$ is plotted for $L=M=100$.  When the interface is close to the central ring, there is an energetic barrier which scales as $2(J_0-J)$ for creating a single crossing point. Expanding this crossing point has no energetic cost. There is, however,
an entropic barrier which is expected to yield a crossing time between the two sides that grows polynomially with $L$ and $M$. The analysis of the average crossing time, $\Delta t$ for several system-sizes is shown \fref{fig:switch_time_eq}, which suggested that $\Delta t \sim L^{4}$.

To conclude, modifying the interaction strength within the central ring was found yield two different wet phases. For $J_0<J$ the interface is strongly pinned to the central ring and fluctuates symmetrically around it, whereas for $J_0>J$ it is repelled from it but crosses it on a time-scale that grows polynomially with $L$ and $M$. In the next section we study the effect of the drive on the interface and find that it leads to a different set of phases.

\section{Analysis of the Ising model with a driven line}
\label{sec:non_eq}
\subsection{Phase diagram for fast drive and slow spin flips on the central ring}
\label{sec:driven_ising}

We now return to the Ising model with a driven line, illustrated in \fref{fig:ising}.
In general the additional Kawasaki dynamics in \eref{eq:ising_drive} leads to a non-equilibrium steady-state, whose form is unknown. However, as explained below, one can derive the phase diagram of the model in the  limit of fast drive and slow spin flips in the central ring.
  The different phases found in this study can be observed numerically also away from this limit, as shown in \sref{sec:numerical}.

  The limit of fast drive is defined such that the relaxation time-scale of the conserving dynamics within
  the central ring is much faster than that of the boundary layers adjacent to that ring. Such a limit allows the conserving dynamics to reach a steady-state between every spin flip in the adjacent boundaries. The steady-state of the conserving dynamics is that of the Asymmetric Simple Exclusion Process (ASEP),
  which is uniform over all the spin-configurations with a fixed number of particles (which corresponds here to a
  a fixed magnetization in the central ring) \cite{derrida1992exact,golinelli2006asymmetric,derrida2007non}. This steady-state is known to be obtained on a time-scale of $\mathcal{O}(L^{3/2})$ \cite{golinelli2004bethe}. On the other hand, since the dynamics in the adjacent rings is non-conserving, their magnetization relaxes on a time-scale of $\mathcal{O}(1)/p_{i,j\neq 0}^{(\mathrm{Gb})}$.

  Similarly, the limit of slow spin flips in the central ring is defined  such that the relaxation time-scale of the non-conserving dynamics in the central ring, given by $\mathcal{O}(1)/p_{i,0}^{(\mathrm{Gb})}$, is much slower than the relaxation time of the adjacent layers. This enables the magnetization in these adjacent layers to reach the steady-state values of the 2D Ising model between every spin flip in the central ring. The limit we consider can thus be written as
\begin{equation}
  p_{i,0}^{(\mathrm{Gb})} \ll p_{i,j\neq 0}^{(\mathrm{Gb})}\ll p^{(\mathrm{Kw})} L^{-3/2}.
\end{equation}

Within these two limits the rate of every spin flip in the central ring is a function only of magnetization in the central ring, defined as
\begin{equation}
m_0(\boldsymbol \sigma)  \equiv  \frac{1}{L} \sum_{i=1}^L\sigma_{i,0}.
\end{equation}
 Summing over the rates of all the possible spin flips in the central ring, one obtains a master equation for $m_0$, which corresponds to that of one-dimensional random walk in a potential. This can be formally written as
\begin{equation}
\label{eq:m0_master}
\fl \frac{d P(m_0)}{dt }  = W_{+}(m_0-\frac{2}{L}) P(m_0-\frac{2}{L})+ W_{-}(m_0+\frac{2}{L}) P(m_0+\frac{2}{L}) - [ W_{+}(m_0)+W_{-}(m_0)] P(m_0),
\end{equation}
where $W_{\pm}$ are the rates of spin flips in the central ring for a given value of $m_0$. These rates can be computed from the known steady-state properties of the 2D Ising model and of the ASEP, as discussed below.  A similar approach has been employed in \cite{Sadhu2012}, where the rates were computed using a low temperature expansion of the Ising model.
A finite temperature analysis was not possible in this case because the properties of the Ising model with a fixed magnetization (or equivalently fixed bulk magnetic field) are unknown.
\Eref{eq:m0_master} has a simple solution of the form
\begin{equation}
P(m_0)=\prod_{X=-L}^{L m_0-2} \frac{W_{+}(X/L)}{W_{-}(X/L+2/L)},
\end{equation}
where $X$ increases in steps of $2$. This can be written in the large $L$ limit as
\begin{equation}
\label{eq:ldf_gen_sol}
P(m_0)\propto e^{\frac{L}{2}\int_{-1}^{m_0} dx \{ \log[ W_{+}(x)/W_{-}(x)] \} + \mathcal{O}(1)}\equiv e^{L\mathcal{G}(m_0) + \mathcal{O}(1)},
\end{equation}
where $\mathcal{G}(m_0)$ is the large deviation function of $m_0$.

\begin{figure}
\noindent
\begin{centering}\includegraphics[scale=0.6]{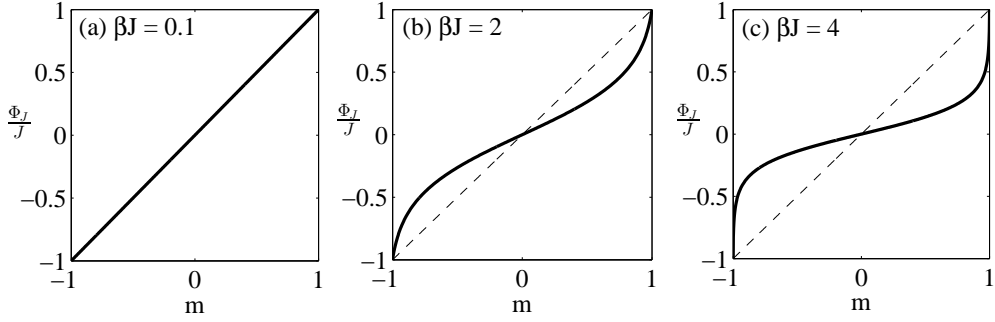}\par\end{centering}
\caption{
\label{fig:hmeff} Effective field induced between the driven line and the boundary line of the two-dimensional Ising model normalized by $J$, $\Phi_J/J$, for different values of $\beta J$.
}
\end{figure}

\Eref{eq:ldf_gen_sol} implies that $\mathcal{G}(m_0)$ depends only on the ratio between $W_{+}(x)$ and $W_{-}(x)$. The first step in obtaining this ratio is to analyze the interaction between line $0$ and its neighbouring lines from above and below. Naively one would think
that for a given magnetization, $m_0$, the spins in line $1$ feel a magnetic field of strength $J m_0$. However, this turns out to be true only for high temperature, $\beta\to 0$.
In order to verify this, one has to consider the ratio between the rate of a spin flip in site $(i,1)$ and the that of the reverse process, given by
\begin{eqnarray}
\fl \qquad \frac{W_{\sigma_{i,1}=-1\to \sigma_{i,1}=1}(m_0)}{W_{\sigma_{i,1}=1\to \sigma_{i,1}=-1}(m_0+2/L)} = \\
\fl \qquad \qquad \frac{[P(\sigma_{i,0}=1|m_0)e^{\beta J }+ P(\sigma_{i,0}=-1|m_0)e^{-\beta J}] e^{\beta J (\sigma_{i-1,1}+\sigma_{i+1,1}+\sigma_{i,2})}}
{[P(\sigma_{i,0}=1|m_0)e^{-\beta J }+ P(\sigma_{i,0}=-1|m_0)e^{\beta J}]e^{-\beta J (\sigma_{i-1,1}+\sigma_{i+1,1}+\sigma_{i,2})}+\mathcal{O}(1/L)}. \nonumber
\end{eqnarray}
The fact that the steady-state measure of the central ring is uniform over all configuration implies that $P(\sigma_{i,0}=\pm 1|m_0) = (1\pm m_0)/2$, which yields
\begin{eqnarray}
\label{eq:effective_h}
\fl  \frac{W_{\sigma_{i,1}=-1\to \sigma_{i,1}=1}(m_0)}{W_{\sigma_{i,1}=1\to \sigma_{i,1}=-1}(m_0+2/L)} &=& \frac{\frac{1+m_0}{2}e^{\beta J \sigma_{i,1}}+ \frac{1-m_0}{2}e^{-\beta J}}
{\frac{1+m_0}{2}e^{-\beta J }+ \frac{1-m_0}{2}e^{\beta J }}
e^{2\beta J (\sigma_{i-1,1}+\sigma_{i+1,1}+\sigma_{i,2})} +\mathcal{O}(\frac{1}{L}) \\
&\equiv &
e^{2\beta [ \Phi_J(m_0) + J (\sigma_{i-1,1}+\sigma_{i+1,1}+\sigma_{i,2})] } +\mathcal{O}(\frac{1}{L}), \nonumber
\end{eqnarray}
where
\begin{eqnarray}
\Phi_J(m)=
-\frac{J}{2}\log \big[\frac{\cosh(\beta J) -m \sinh(\beta J))}{\cosh(\beta J) +m \sinh(\beta J)}\big].
\end{eqnarray}
The spins in line $1$ therefore feel an effective magnetic field $\Phi_J(m_0)$ which is in general different than $J m_0$. This field is plotted
for various values of $\beta J$ in \fref{fig:hmeff}. \Fref{fig:hmeff}a demonstrates the fact that $\lim_{\beta\to 0} \Phi_J(m)= J m$.
The fact that $\Phi_J(m)\neq J m$ was noted also in \cite{hilhorst2011two}, where the Ising interaction between two lines of spins was analyzed in the case where one line
is strongly driven.

Since the spin flips
in the driven line occur slowly, the systems above and below the driven line relax between every such spin flip to the steady-state of the
2D Ising model with boundary fields $h_\downarrow, \Phi_J(m_0)$ and $\Phi_J(m_0),h_\uparrow$, respectively \footnote{The fact that the driven line exerts
a field of strength $\Phi_J(m_0)$ and not $Jm_0$ was unaccounted for in \cite{Sadhu2012}.
However, by repeating the derivation presented there, we verified that the use of $\Phi_J$ does
 not change the qualitative results of that study.}. Consequently, the magnetization in lines $1$ and $-1$ is given on the time-scale in which $m_0$ is fixed by
 \begin{equation}
 \label{eq:m1_mm1}
 m_1=\mathcal{M}(\Phi_J(m_0),h_\uparrow), \qquad m_{-1}=\mathcal{M}(\Phi_J(m_0),h_\downarrow).
 \end{equation}

Using an analysis similar to that of \eref{eq:effective_h}, the ratio between the rates of spin flips within the driven line are given by
\begin{equation}
\label{eq:spin0flip}
\frac{W_{\sigma_{i,0}=-1\to \sigma_{i,0}=1}(m_0)}{W_{\sigma_{i,0}=1\to \sigma_{i,0}=-1}(m_0+2/L)}
 = e^{2\beta[ 2 \Phi_{J_0}(m_0)+ \Phi_{J}(m_1)+\Phi_{J}(m_{-1})]} +\mathcal{O}(\frac{1}{L}).
\end{equation}
This implies that apart for the average magnetic field from the two adjacent rings, $\Phi_{J}(m_{-1})$ and $\Phi_{J}(m_{1})$, every spin in the central ring feels a field of strength $2\Phi_{J_0}(m_0)$, exerted on it by the spins within the central ring.

The fact that the expression in \eref{eq:spin0flip} is independent of $i$, enables us to derive from it the ratio between the rates of increase and decrease of $m_0$ by summing
over all possible single spin flips, yielding
\begin{eqnarray}
\frac{W_{+}(m_0)}{W_{-}(m_0+2/L)} &=&
\frac{\sum_{i=1}^L \delta_{\sigma_{i,0},+1} W_{\sigma_{i,0}=-1\to \sigma_{i,0}=1}(m_0)}{
\sum_{i=1}^L \delta_{\sigma_{i,0},+1} W_{\sigma_{i,0}=1\to \sigma_{i,0}=-1}(m_0+2/L)}  \\
&=& \frac{L(1+m_0)/2}{L(1-m_0)/2}e^{2\beta[ 2 \Phi_{J_0}(m_0)+ \Phi_{J}(m_1)+\Phi_{J}(m_{-1})]} +\mathcal{O}(\frac{1}{L}). \nonumber
\end{eqnarray}
Inserting the expression for $m_{\pm1}$ in \eref{eq:m1_mm1} into the above equation yields an expression for $W_{+}(m_0)/W_{-}(m_0)$ which is
fully determined by $m_0$ and is given by
\begin{equation}
\fl \qquad \frac{W_{+}(m_0)}{W_{-}(m_0)} = \frac{(1+m_0)}{(1-m_0)}e^{2\beta[ 2 \Phi_{J_0}(m_0)+ \Phi_{J}\big(  \mathcal{M}(\Phi_J(m_0),h_\uparrow) \big)
+\Phi_{J}\big(  \mathcal{M}(\Phi_J(m_0),h_\downarrow) \big)]} +\mathcal{O}(\frac{1}{L}).
\end{equation}
Finally, the above result can be used in the solution of the large deviation function in \eref{eq:ldf_gen_sol}, yielding
\begin{eqnarray}
\label{eq:Gm0final}
\fl \qquad \mathcal{G}(m_0)&=&-\frac{1+m_0}{2}\log(\frac{1+m_0}{2})-\frac{1-m_0}{2}\log(\frac{1-m_0}{2})+2\beta\int^{m_0} dx \Phi_{J_0}(x) \\
\fl      &&+\beta \int^{m_0} dx \Big [ \Phi_J\big ( \mathcal{M}(\Phi_J(m_0),h_\uparrow) \big) +  \Phi_J\big ( \mathcal{M}(\Phi_J(m_0),h_\downarrow) \big) \Big ]. \nonumber
\end{eqnarray}
Note that the first two terms in the RHS above correspond to the entropy of the central ring. The following two integrals account for the interaction within the central ring,
and with the two adjacent rings, respectively.

\begin{figure}
\noindent
\begin{centering}\includegraphics[scale=0.6]{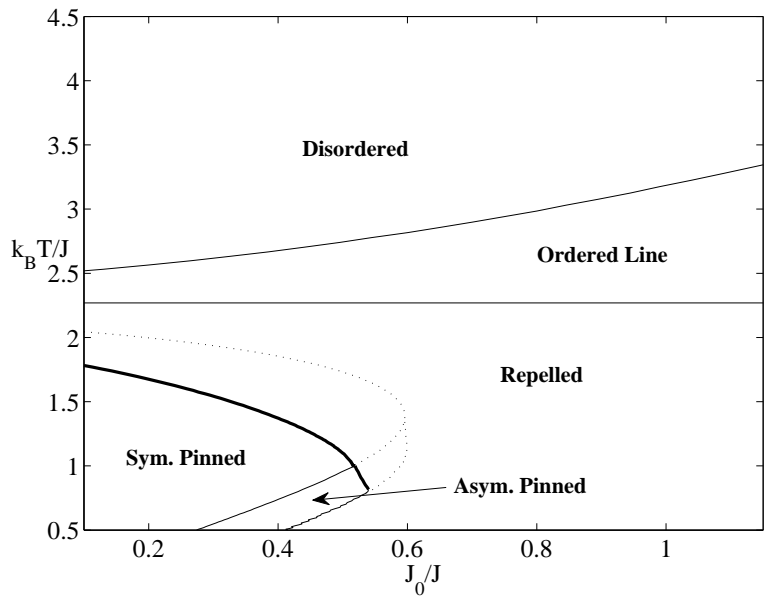}\par\end{centering}
\caption{
\label{fig:ising_phase}
Phase diagram of the 2D Ising model with a driven line in the limit of strong drive and slow spin flips on the driven line. The
boundary fields are such that prohibit a dry state, i.e. $h_\downarrow<-J$ and $h_\uparrow>J$.
First and second order transition line are denoted by thick and thin solid lines, respectively. The dotted line denotes the stability limits of the two pinned phases. Here the boundary fields are given by $h_\downarrow <J$ and $h_\uparrow>J$, such that the system is always in the wet phase. Their exact values do not affect the results in this case.
}
\end{figure}
\begin{figure}
\noindent
\begin{centering}\includegraphics[scale=0.6]{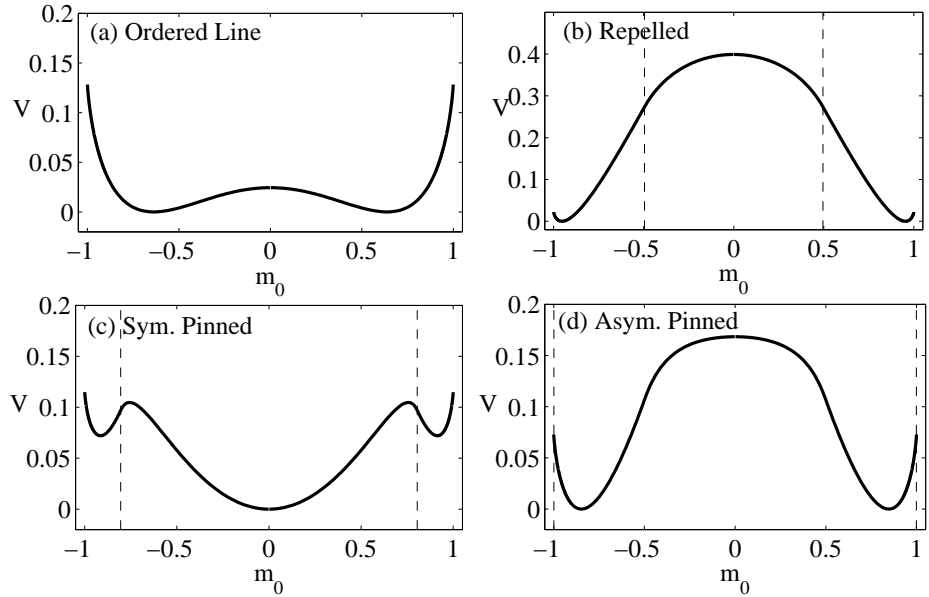}\par\end{centering}
\caption{
\label{fig:ising_V}
The effective potential for the random walk of $m_0$, $V(m_0)=-\mathcal{G}(m_0)$, computed for the following points in the phase diagram in \fref{fig:ising_phase}:
(a) $\beta J =0.35, J_0/J=1$, (b) $\beta J =0.5, J_0/J=1$ (c) $\beta J =0.66, J_0/J=0.2$, (d) $\beta J =2, J_0/J=0.4$.
The type of phase depicted in each figure is denoted in the corresponding caption. The dashed lines denote the value of $m_0$ for which
$\Phi_J(m_0)=\pm h_w(T)$. For $|\Phi_J(m_0)|>|h_w(T)|$ the interface is detached from the boundary. Here the boundary fields are given by $h_\downarrow <J$ and $h_\uparrow>J$, such that the system is always in the wet phase. Their exact values do not affect the results in this case.  }
\end{figure}

By studying the extrema of $\mathcal{G}(m_0)$ for different values of $\beta$ and $J_0/J$ one can obtain the phase diagram of the model.
The resulting phase diagram is plotted in \fref{fig:ising_phase} for the relatively simple case where $h_\downarrow<-J, h_\uparrow>J$, in which
 the interface is never attached to the boundaries of the cylinder. Representing plots of $\mathcal{G}(m_0)$ in each phase are provided in \fref{fig:ising_V}.

The main features of the phase diagram are as follows: In the driven model, the central line become magnetized before the bulk of the system does, i.e. for  $T>T_c \approx 2.269J/k_B$.
The system undergoes a second order phase transition between two phases where the bulk of the system is disordered and the central ring is either paramagnetic, $m_0= 0$, or ferromagnetic ,$m_0\neq 0$
. The plot of $\mathcal{G}(m_0)$ for this phase, referred here as the {\it ordered line} phase, is shown in \fref{fig:ising_V}a. Below the Ising critical point and
for $J=J_0$ the system exhibits only one type of a wet phase, where the interface is repelled from the driven line, as shown in \fref{fig:ising_V}b. The sign-flip symmetry of $m_0$ is spontaneously broken in this phase, yielding two degenerate states, where the interface is either above of below the driven line. As suggested by the form $\mathcal{G}(m_0)$, there is a barrier between those two degenerate states which increases with $L$. This implies that the tunneling time between them increases exponentially with $L$, as confirmed numerically in the next section. This is in contrast to the polynomial growth of the crossing-time found in the equilibrium repelled phase in \sref{sec:interface}. In this respect the repelled phase observed here can be considered to be a `strongly-repelled' phase.

As in equilibrium, decreasing $J_0$ below $J$ is generally expected to attract the interface to the driven line. However, in this case the effective repulsion due to the drive implies that one has to go significantly below $J_0=J$ in order to observe a pinned phase. For $J_0 \lesssim 0.54 J$, one finds two phases
where the interface is pinned to the central ring. For relatively high temperatures, we observe a phase similar to that observed in equilibrium, where the interface fluctuates symmetrically around the central ring, yielding $m_0=0$ (corresponding LDF shown in \fref{fig:ising_V}c). The transition between this phase and the disordered phase is of first order (thick solid line in \fref{fig:ising_phase}).
 As the temperature is decreased the sign-flip symmetry of $m_0$ is spontaneously broken, yielding a phase where
 the interface found in one of the sides of the central ring but at a distance from it that does not scale with $L$. The transition between the {asymmetrically-pinned phase} and the {symmetrically-pinned phase} is of second order.
The plot of $\mathcal{G}(m_0)$ in the {asymmetrically-pinned phase} is shown in \fref{fig:ising_V}d.
A reentrant second order transition back into the repelled phase
is observed at lower temperature, where the field induced by the driven line is strong enough to unpin the interface, namely $|\Phi_J(m_0)|>|h_w(T)|$.

The asymmetrically-pinned phase was observed in \cite{Sadhu2012}, where Kawasaki dynamics was considered in the bulk and a low temperature limit was taken. The finite
temperature fluctuation of the interface in this phase can thus be understood using the known boundary properties of the two-dimensional Ising model, analyzed extensively
in \cite{McCoy}.

\subsection{Numerical results for the driven case}
\label{sec:numerical}
The phase diagram of the model presented in \sref{sec:model} was computed in \sref{sec:driven_ising} in the limit of $M \sim L\to\infty$, fast drive
and slow spin flips on the central ring, $p_{i,0}^{(\mathrm{Gb})} \ll p_{i,j\neq 0}^{(\mathrm{Gb})}\ll p^{(\mathrm{Kw})} L^{-3/2}$. In this section we test numerically whether the results obtained in that limit yield a correct qualitative
picture  for finite $L,p^{\mathrm{Gb}}_{i,j}$ and $p^{\mathrm{Kw}}$. For simplicity we consider here $M=L$. The results appear not to depend much on the ratio between $p^{\mathrm{Gb}}_{i,0}$ and $p^{\mathrm{Gb}}_{i,j\neq 0}$ and the asymmetry $q$ or the drive. We thus choose to study here the simple case where  $p^{\mathrm{Gb}}_{i,j}=1$ for all $j$'s and  $q=1$ (totally asymmetric Kawasaki dynamics). On the other hand we find very different results for large and small $p^{\mathrm{Kw}}$. To test the prediction of the above analysis we considered relatively fast drive by taking $p^{\mathrm{Kw}}=1000$.

\begin{figure}
\noindent
\begin{centering}\includegraphics[scale=0.7]{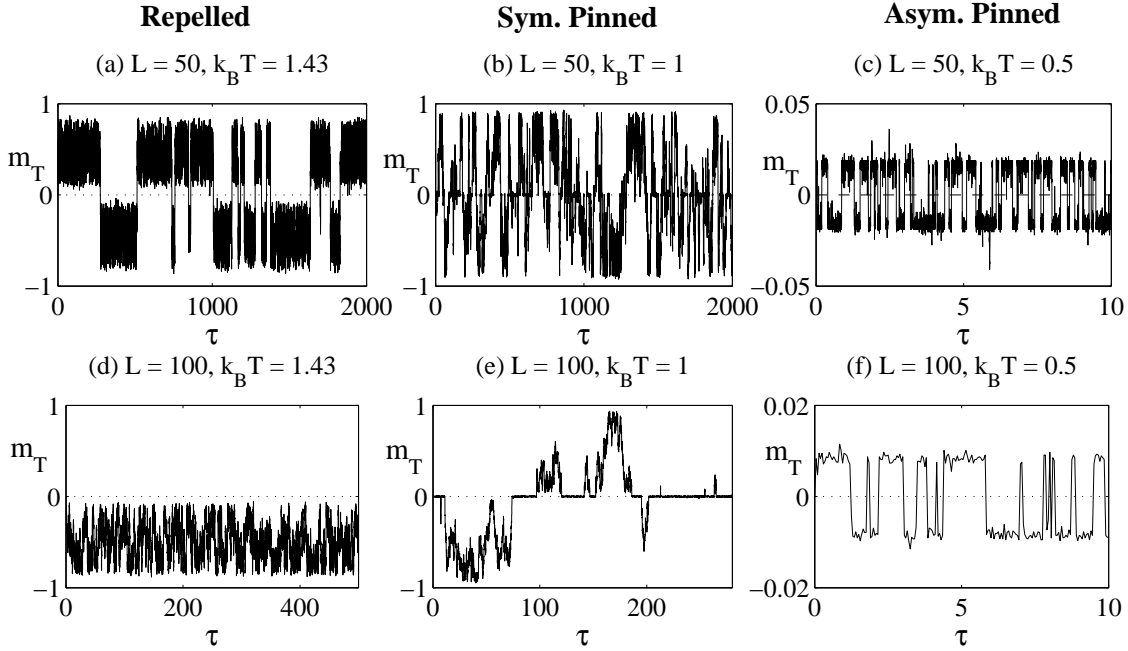}\par\end{centering}
\caption{\label{fig:mt}
  The overall magnetization,  $m_t(\boldsymbol \sigma)$,
 measured in simulation as a function of $\tau = t/L^2$ for $p^{(\mathrm{Kw})}=1000$, $J_0=J$, $M=L$ and $L=50,100$.
 The values of $L$ and $\beta$ used is mentioned in the caption of each figure.
 The three columns represent the three different ordered phases observed in the theoretical calculation in \fref{fig:ising_phase}.
 }
\end{figure}

\begin{figure}
\noindent
\begin{centering}\includegraphics[scale=0.7]{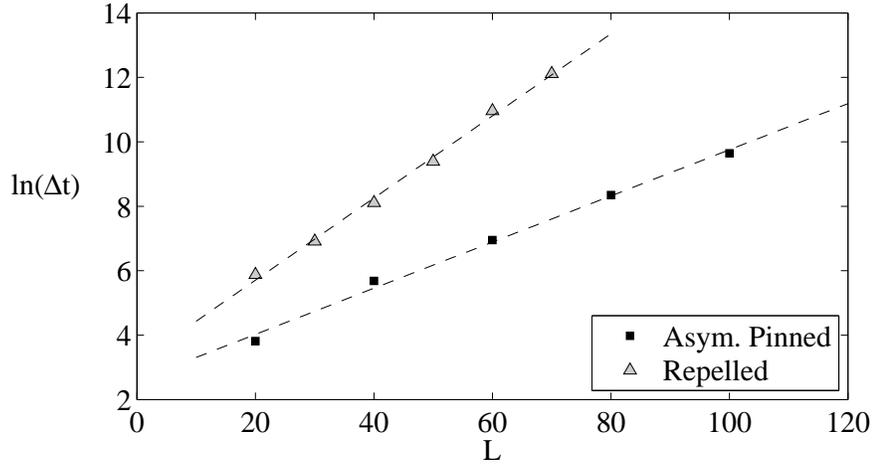}\par\end{centering}
\caption{\label{fig:switch_time_driven}
The logarithm of the transition time between the two degenerate ground state, $\ln (\Delta t)$, as a function of the system width, $L$, as measured for
$L=M$, $p^{(\mathrm{Gb})}_{i,j}=1$ and $p^{(\mathrm{Kw})}=1000$. The hollow triangles ($\bigtriangleup$) represent the repelled phase, observed for $k_BT=1.43$, whereas the squares ($\blacksquare$) represents the asymmetrically-pinned phase, observed at $k_BT=0.5$. The dashed lines show the corresponding linear fits, with a slope of 0.072 in the pinned phase and a slope of 0.127 in the repelled phase.
 }
\end{figure}

We focus here on the three different wet phases, which appear for $J_0 \lesssim 0.54J$. We consider specifically $J_0 = 0.4J$. The three phases are clearly visible when observing the time evolution of the overall magnetization, defined as
\begin{equation}
m_T(\boldsymbol \sigma) \equiv  \frac{1}{L(2M+1)} \sum_{i=1}^L\sum_{j=-M}^M\sigma_{i,j}.
\end{equation}
Its behaviour in the repelled, symmetrically-pinned and asymmetrically-pinned phases is plotted respectively in \figs \ref{fig:mt}a,\ref{fig:mt}b and \ref{fig:mt}c for $L=50$ and in  \figs \ref{fig:mt}d,\ref{fig:mt}e and \ref{fig:mt}f for $L=100$. In the repelled phase, the system can be in two degenerate state where the magnetization is significantly different from $0$. The transitions
between these two states are clearly visible for $L=50$, but cannot be observed for $L=100$ due to the limited simulation time. The exponential growth of the transition time is depicted by the grey triangles in \fref{fig:switch_time_driven}, which shows the logarithm of the average time interval between the sign-flips of $m_T(\boldsymbol \sigma)$ as function of $L$. The symmetrically-pinned phase, where $\langle m_T\rangle =0$, appears in \figs \ref{fig:mt}b and \ref{fig:mt}e to become more stable as the system-size increases. For system of size $L=200$ we were unable to observe the detachment of the interface.
In the asymmetrically-pinned phase, shown in \figs \ref{fig:mt}c and \ref{fig:mt}f, the system moves between two states where $m_T \sim \pm \frac{1}{M}$. As shown by the squares in \fref{fig:switch_time_driven}, the crossing-time of the interface between those states grows exponentially with $L$ but at a smaller rate than that of the repelled phase.

\begin{figure}
\noindent
\begin{centering}\includegraphics[scale=0.65]{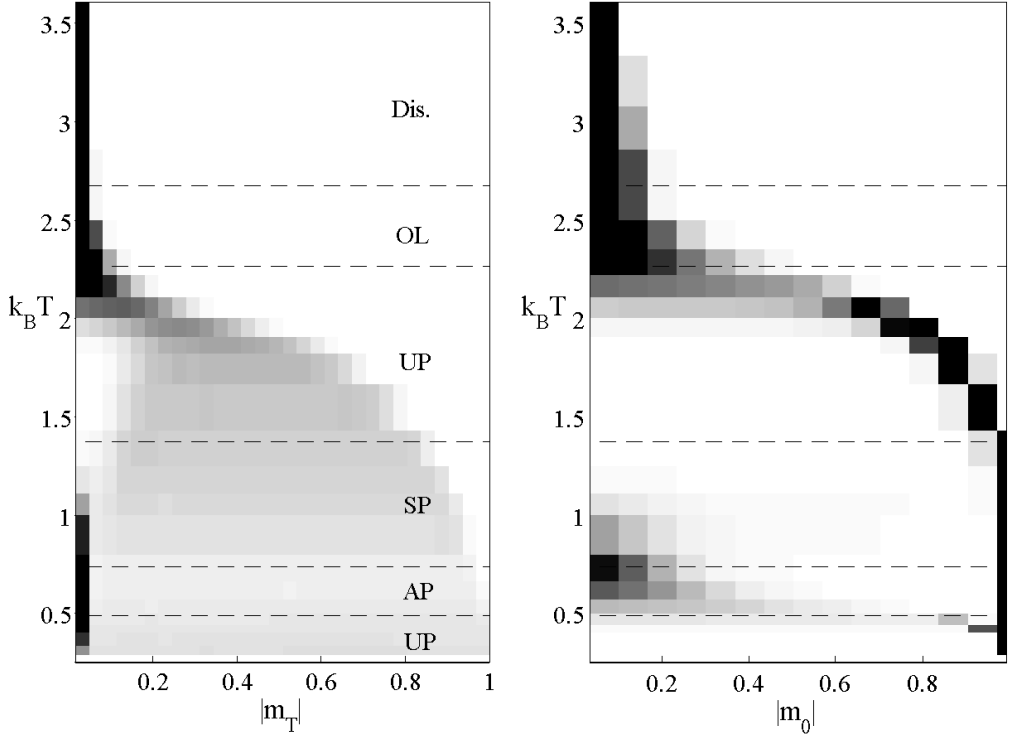}\par\end{centering}
\caption{\label{fig:histogram1}
Histogram of the total magnetization, $|m_T|$, and the magnetization on the central ring, $|m_0|$, as a function of the temperature.
The data was measured for $L=M=30$, $p^{(\mathrm{Gb})}_{i,j}=1$ and $p^{(\mathrm{Kw})}=1000$. Darker colours correspond for higher probability.
The different phases as obtained in the limit of fast driven and slow spins flips on the central rings as denoted on the graph (OL for ordered line, R for repelled, SP for symmetrically-pinned
AP for asymmetrically-pinned). The histogram of $m_T$ is smoothen over to a scale of $1/M$.
  }
\end{figure}

\begin{figure}
\noindent
\begin{centering}\includegraphics[scale=1.25]{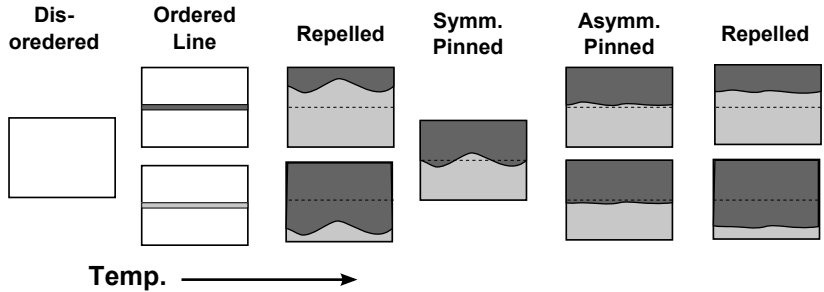}\par\end{centering}
\caption{\label{fig:phases}
Sketch of the different phases shown in \fref{fig:histogram1}.  The state of the system is sketched in boxes that signify a typical spin state in the cylinder, where
light, dark gray and white denote the plus phase, minus phases and disordered phase, respectively.
  }
\end{figure}

The transitions between the various phases are examined in \fref{fig:histogram1}, which shows the histogram of the absolute value of the overall magnetization, $|m_T(\boldsymbol \sigma)|$,
and of the magnetization on the driven line, $|m_0(\boldsymbol \sigma)|$.
 The histogram is measured for a relatively small system size, $L=M=30$, for which the transition between the different metastable states could be observed.
Above the Ising critical temperature ($T_c\approx 2.269k_BT$) one finds a smooth transition into the ordered-line phase (OL) where $|m_0|$ increases $0$ whereas $|m_T|$ is still relatively close to $0$. Below it we find the repelled phase (R) where both $|m_0|$ and $|m_T|$ are large. At even lower temperatures there a coexistence of the repelled phase and the symmetrically-pinned (SP) phase,
where $|m_0|\approx 0$ and $m_T\approx0$. As the temperature is further decreased, the pinned phase turns into an asymmetrically-pinned (AP) where $|m_0|>0$ and $m_T\sim \frac{1}{M}$. Finally at very low temperatures we find evidence for a reentrant transition into the repelled phase. For convenience, the state of the system at each phase is sketched in \fref{fig:phases}.
These phases appear more or less at the temperatures calculated for fast drive and slow spin flips on the driven line. These theoretical transition points are denoted by the dashed lines in \fref{fig:histogram1}.

\section{Conclusions}
 \label{sec:conc}
We studied the Ising model on a cylinder with Glauber dynamics in the bulk and biased Kawasaki dynamics in the central ring of the cylinder. The purpose of this study
is to analyze the effect of a localized drive on the properties of an interface between two coexisting phases.

 In order to better understand the effect of the drive we considered a modified interaction strength within the central ring, $J_0$, as well as analyzed the properties of the corresponding equilibrium model (where the drive is turned off). The equilibrium model has been studied previously for $J_0<J$, using an approximate RSOS description. The model exhibits in this case a symmetrically-pinned phase, where the interface is pinned to the central ring and fluctuates symmetrically around it. We complemented this study by considering the case of $J_0>J$, where the interface is found to be repelled from the central ring.

The driven model is analyzed in a special limit of fast drive and slow spin flips on the central ring. Within this limit, the large deviation of the magnetization on the central ring, $m_0$, can be computed and the phase diagram of the model can be drawn. The driving field does not alter the critical temperature in this limit.
Above the Ising critical temperature, the system exhibits a second order phase transition between a phase where the bulk is disordered and $m_0=0$, and a phase where the bulk is disordered but $m_0\neq0$. Below the critical temperature the system exhibits three different ordered phases: a repelled phase, where the interfaces is at a macroscopic distance from the central ring, a symmetrically-pinned phase, with $m_0=0$, and an asymmetrically-pinned phase, where the interface is at a microscopic distance from the central ring and thus $m_0\neq0$.

The repelled phase is found to be more stable in the driven model in the sense that the time it takes interface to cross the central ring increases exponentially with the circumference of the cylinder, $L$. This is in contrast with the equilibrium model with $J_0>J$, where the crossing-time grows polynomially with $L$ and the height of the cylinder, $M$. The strong repulsion of the interface in the driven case appears to be due to long-range correlation induced by the drive.

The asymmetrically-pinned phase, which was found only in the driven model, was observed also in the magnetization-conserving variant of our model in \cite{Sadhu2012}. In this case the model can be studied analytically in the limit of strong drive,
slow spin-exchanges with the central ring and low temperature. In the low temperature limit the authors have found only asymmetrically-pinned phase, in contrast to the repelled phase we find in the non-conserving model. In the present study we are able go beyond the low temperature limit and draw the full phase diagram of the model. The finite temperature analysis gives a better insight into the fluctuations of the interface in the asymmetrically-pinned phase.

We test our observation numerically by simulating a finite system. When considering relatively fast drive, one finds the evidence for the different phases found  analytically in the limit of fast drive and slow spin flips in the central ring.

\ack We thank A. Bar, O. Hirschberg and T. Sadhu for helpful
discussions. The support of the Israel Science Foundation (ISF) and of the Minerva Foundation with
funding from the Federal German Ministry for Education and Research is gratefully
acknowledged. We also thank the Galileo Galilei Institute for Theoretical Physics for
the hospitality and the INFN for partial support during the completion of this work.

\appendix

\section{Solution of the Ising model on a cylinder with two boundary fields}
\label{sec:exact}
In this appendix we present a generalization of the solution of the Ising model on a cylinder, derived in \cite{McCoy} for the case of a single boundary field, to the
case two boundary fields. Since the two derivations are relatively similar, we refrain from repeating many of the details presented in \cite{McCoy} and simply emphasize the differences
between the two cases.

\begin{figure}
\noindent
\begin{centering}\includegraphics[scale=1.3]{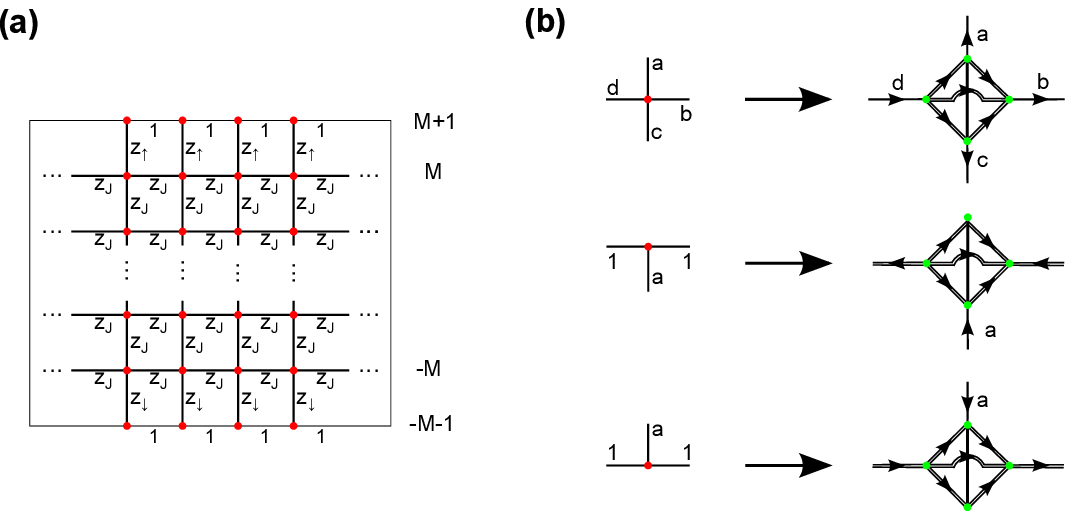}\par\end{centering}
\caption{
\label{fig:ising_graph}
Inset (a): the square lattice used in the partition sum in \eref{eq:app_Z2} is computed. The sum is over all the spanning graphs with weight of each graph is the product of edge weights, denoted in the graph. Inset (b): the mapping of each vertex in (a) into a four-vector, used in \eref{eq:app_detA}. }
\end{figure}

The partition function of the Ising model is given
\begin{equation}
Z=\sum_{\boldsymbol \sigma} e^{-\beta H[\boldsymbol \sigma]},
\end{equation}
where the Hamiltonian is defined for a system of size $L\times (2M+1)$ as
\begin{equation}
H[\boldsymbol \sigma]=-J\sum_{i=1}^L \sum_{j=-M}^{M-1} ( \sigma_{i,j}\sigma_{i,j+1} + \sigma_{i,j}\sigma_{i+1,j} )
+\sum_{i=1}^L ( h_{\downarrow} \sigma_{i,-M} + h_{\uparrow} \sigma_{i,M}).
\end{equation}
The first step in the computation of $Z$ is to write it in the following form:
\begin{eqnarray}
\label{eq:ap_Z1}
\fl\qquad Z &=&  \cosh^{ 4 L M }(\beta J) \cosh^L(\beta h_\uparrow) \cosh^L(\beta h_\downarrow)\\
\fl\qquad  &&\times\sum_{\boldsymbol \sigma} \prod_{i=1}^L \prod_{j=-M}^{M-1} [  (1+z_J \sigma_{i,j}\sigma_{i,j+1})(1+ \sigma_{i,j}\sigma_{i+1,j} )]
\prod_{i=1}^L [ ( 1+z_\downarrow \sigma_{i,-M})(1+z_\uparrow\sigma_{i,M})].\nonumber
\end{eqnarray}
where
\begin{equation}
z_J=\tanh(\beta J), \qquad z_\downarrow = \tanh(\beta h_\downarrow), \qquad z_\uparrow = \tanh(\beta h_\uparrow).
\end{equation}
The sum in \eref{eq:ap_Z1} can be expressed in a graphical form as
\begin{equation}
\label{eq:ap_Z2}
Z =  2^{L(2M+1)} \cosh^{ 4 L M }(\beta J) \cosh^L(\beta h_\uparrow) \cosh^L(\beta h_\downarrow)
\sum_{p,q,r} z_J^p z_\downarrow^q z_\downarrow^r N_{p,q,r},
\end{equation}
where $N_{p,q,r}$ is the number of figures that can be drawn on the lattice in \fref{fig:ising_graph}a with $p$ inner bonds (denoted by $z_J$),
$q$ and $r$ bonds connecting to the upper and lower lines respectively (denoted by $z_\downarrow$ and $z_\uparrow$, respectively) and with every bond used at most once.

Following the derivation in \cite{McCoy} the sum in \eref{eq:ap_Z2} can be performed by counting all the spanning graphs in a corresponding directed graph obtained as follows:
Every vertex in \fref{fig:ising_graph}a is mapped into four vertices according to \fref{fig:ising_graph}b, where $a,b,c,d$ denote a
 general number which can be either $z_J,z_\downarrow$ or $z_\uparrow$. Single directed bonds in \fref{fig:ising_graph}b can be drawn only in the stated direction and their weight is written above them. The weight of the double bonds in \fref{fig:ising_graph}b is $1$ when drawn in the direction of the arrow and $-1$ otherwise. The weights of the directed graph
  is the product of weights of all the bonds. This directed graph can be repressed by an adjacency matrix, $A$, whose dimension is $4L(2M+3)\times 4L(2M+3)$ and
  whose entries contain the weight of the bonds.
  For the sake of brevity we refrain from writing the form of the matrix explicitly. Its form for the case of a single boundary field can be found in Eq. VI.2.6 in \cite{McCoy}.

   As discussed in \cite{McCoy}, the partition function can be written in terms of the Pfaffian of the adjacency matrix as
\begin{equation}
Z=2^{L(2M+1)-1} \cosh^{4ML}(\beta J)\cosh^L(\beta h_\uparrow)\cosh^L(\beta h_\downarrow) \mathrm{Pf} A.
\end{equation}
Using the properties of the Pfaffian operator, whereby $(\mathrm{Pf} A)^2 = \det A$ (proven in Sec. IV.2 of \cite{McCoy}), one can write the partition function as
\begin{equation}
\label{eq:app_Z2}
Z^2=2^{2L(2M+1)-2} \cosh^{8ML}(\beta J)\cosh^{2L}(\beta h_\uparrow)\cosh^{2L}(\beta h_\downarrow) \det A.
\end{equation}

In \cite{McCoy} the fact that $A$ is nearly cyclic in the horizontal direction is used in order to write $\det A$ as a product over matrices that represent only a vertical slice of the
lattice. This transformation and an additional one, where the left and right sites in every quadruplet in \fref{fig:ising_graph}b
  are removed as well, can be performed here by following the derivation
presented in  Sec. VI.3 of \cite{McCoy}. Due to their length we omit the algebraic manipulation and write down the final result:
\begin{equation}
\label{eq:app_detA}
\det A = \prod_{n=1}^{L} \big[ |1+e^{i\theta_n}|^2 |1+z_J|^{2M+1} \det \varrho(\theta_n) ]
\end{equation}
where
\begin{equation}
\theta_n = \frac{\pi (2n-1)}{L}.
\end{equation}
Here $\varrho(\theta)$ denotes a $(4M+4)\times (4M+4)$ matrix given by
\begin{equation}
\label{eq:rho_mat}
\varrho(\theta)= \left(\begin{array}{cccccccccccc}
-c & 0\\
0 & c & z_{\downarrow}\\
 & -z_{\downarrow} & -a & b\\
 &  & -b & a & z_{J}\\
 &  &  & z_{J} & -a & b\\
 &  &  &  & -b & a & z_{J}\\
 &  &  &  &  & -z_{J} & . & .\\
 &  &  &  &  &  & . & . & .\\
 &  &  &  &  &  &  & . & -a & b\\
 &  &  &  &  &  &  &  & b & a & -z_{\uparrow}\\
 &  &  &  &  &  &  &  &  & z_{\uparrow} & c & 0\\
 &  &  &  &  &  &  &  &  &  & 0 & -c
\end{array}\right), \quad
\end{equation}
where
\begin{equation}
a = \frac{2i z_J \sin \theta}{|1+z_J e^{i\theta}|^2}, \qquad b = \frac{1-z_J^2}{|1+z_J e^{i\theta}|^2}, \qquad c = \frac{2i \sin \theta}{|1+z_J e^{i\theta}|^2}.
\end{equation}
Each pair of row and column represent the up and down sites in a single vertical slice of the directed graph.
In \cite{McCoy} the same form of $\varrho(\theta)$ was obtained, excluding the last two rows and columns which result in the present case from the interaction of the spins with the lower boundary field.

In the following we evaluate $\det \varrho(\theta)$ and demonstrate that the additional rows and columns in \eref{eq:rho_mat} yield the free energy term of the lower
 boundary which is identical to that obtained in \cite{McCoy} for case of a the single boundary field.
Computing the determinant of tri-diagonal matrices of the form,
\begin{equation}
f_{n}=\left|
\begin{array}{ccccc}
a_{1} & b_{1}\\
c_{1} & a_{2} & b_{2}\\
 & c_{2} & \ddots & \ddots\\
 &  & \ddots & \ddots & b_{n-1}\\
 &  &  & c_{n-1} & a_{n}
\end{array}
\right|,
\end{equation}
can be done by the following recursive relation:
\begin{equation}
f_{n}=a_{n}f_{n-1}-c_{n-1}b_{n-1}f_{n-2},
\end{equation}
where $f_{0}=1$ and $f_{-1}=0$. Since in the bulk of the matrix $\varrho(\theta)$ repeats itself every two rows, i.e. $a_{n+2}=a_n, b_{n+2}=b_n,c_{n+2}=c_n$,
 it is useful to express the recursive relations of $f_n$ and $f_{n-1}$ as
\begin{equation}
\label{eq:app_rec_mat}
\left(\begin{array}{c}
f_{n}\\
f_{n-1}
\end{array}\right)=\left(\begin{array}{cc}
a_{n}a_{n-1}-c_{n-1}b_{n-1} & -a_{n}c_{n-2}b_{n-2}\\
a_{n-1} & -c_{n-2}b_{n-2}
\end{array}\right)\left(\begin{array}{c}
f_{n-2}\\
f_{n-3}
\end{array}\right)
\end{equation}
For the analysis of $\det \varrho(\theta)$ we denoted by $\mathcal{R}_n$ and $\mathcal{D}_n$ the determinants of the sub-matrices of $\varrho(\theta)$ with the first $2n+2$ and $2n+1$ identical to those in   \eref{eq:rho_mat}, respectively.

  \Eqs (\ref{eq:rho_mat}) and (\ref{eq:app_rec_mat}) imply that for $n>1$
\begin{equation}
\label{eq:ap_rec1}
\left(\begin{array}{c}
\mathcal{R}_{n}\\
z_J\mathcal{D}_{n}
\end{array}\right)=\left(\begin{array}{cc}
-a^{2}+b^{2} & az_J\\
-az_J & z^{2}_J
\end{array}\right)\left(\begin{array}{c}
\mathcal{D}_{n-1}\\
z_J\mathcal{D}_{n-1}
\end{array}\right).
\end{equation}
For $n=1$ one obtains a different equation, given by
\begin{equation}
\left(\begin{array}{c}
\mathcal{R}_{1}\\
z_J\mathcal{D}_{1}
\end{array}\right)=\left(\begin{array}{cc}
-a^{2}+b^{2} & az_{J}\\
-az_{J} & z_{J}^{2}
\end{array}\right)\left(\begin{array}{c}
\mathcal{R}_{0}\\
z_{J}^{-1}z_{\downarrow}^{2}\mathcal{D}_{0}
\end{array}\right)
\end{equation}
where $\mathcal{R}_{0}=-c^2$ and $\mathcal{D}_{0}=-c$. For reasons that will be clear below it is more useful to express the above equation as
\begin{equation}
\left(\begin{array}{c}
\mathcal{R}_{0}\\
z_{J}^{-1}z_{\downarrow}^{2}\mathcal{D}_{0}
\end{array}\right)=\left(\begin{array}{c}
-c^{2}\\
-z_{J}^{-1}z_{\downarrow}^{2}c
\end{array}\right)=\left(\begin{array}{cc}
-c^{2} & cz_{J}\\
-cz_{J}^{-1}z_{\downarrow}^{2} & z_{\downarrow}^{2}
\end{array}\right)\left(\begin{array}{c}
1\\
0
\end{array}\right).
\end{equation}
The recursion relation for the last two rows and columns of $\varrho(\theta)$ yields its determinant, given by
\begin{equation}
\label{eq:ap_rec3}
\left(\begin{array}{c}
\det\varrho(\theta)\\
z_{J}\det\varrho'(\theta)
\end{array}\right)=\left(\begin{array}{cc}
-c^{2} & -cz^{-1}z_{\uparrow}^{2}\\
cz & z_{\uparrow}^{2}
\end{array}\right)\left(\begin{array}{c}
\mathcal{R}_{2M}\\
z\mathcal{D}_{2M}
\end{array}\right),
\end{equation}
where $\varrho'(\theta)$ is the matrix in \eref{eq:app_rec_mat} with the last row and column removed.
\Eqs (\ref{eq:ap_rec1})-(\ref{eq:ap_rec3}) yield the following expression for $\det \varrho(\theta)$:
\begin{eqnarray}
\fl \quad \det\varrho(\theta)
&=&\left(\begin{array}{c}
1\\
0
\end{array}\right)^{T}\left(\begin{array}{cc}
-c^{2} & -cz^{-1}z_{\uparrow}^{2}\\
cz & z_{\uparrow}^{2}
\end{array}\right)\left(\begin{array}{c}
\mathcal{R}_{2M}\\
z\mathcal{D}_{2M}
\end{array}\right) \\
\fl \quad  &=&\left(\begin{array}{c}
1\\
0
\end{array}\right)^{T}\left(\begin{array}{cc}
-c^{2} & -cz^{-1}z_{\uparrow}^{2}\\
cz_J & z_{\uparrow}^{2}
\end{array}\right)\left(\begin{array}{cc}
-a^{2}+b^{2} & az_{J}\\
-az_{J} & z_{J}^{2}
\end{array}\right)^{2M}\left(\begin{array}{cc}
-c^{2} & cz_{J}\\
-cz_{J}^{-1}z_{\downarrow}^{2} & z_{\downarrow}^{2}
\end{array}\right)\left(\begin{array}{c}
1\\
0
\end{array}\right).  \nonumber
\end{eqnarray}
As in the standard transfer matrix method, the above equation can be expressed in terms of the eigenvalues and eigenvectors of the matrix in \eref{eq:ap_rec1}, denoted by $\lambda_1,\lambda_2$ and ${\bf v}_1$ and ${\bf v}_2$, yielding
\begin{equation}
\label{eq:app_detrho}
\fl \det\varrho(\theta)=\sum_{k=1,2}\left[\left(\begin{array}{c}
1\\
0
\end{array}\right)^{T}\left(\begin{array}{cc}
-c^{2} & -cz^{-1}z_{\uparrow}^{2}\\
cz_J & z_{\uparrow}^{2}
\end{array}\right)v_{k}\right]\lambda^{2M}\left[v_{k}^{T}\left(\begin{array}{cc}
-c^{2} & cz_{J}\\
-cz_{J}^{-1}z_{\downarrow}^{2} & z_{\downarrow}^{2}
\end{array}\right)\left(\begin{array}{c}
1\\
0
\end{array}\right)\right].
\end{equation}

In \cite{McCoy}, due to the absence of the upper magnetic field, $\det\varrho(\theta)$ is given by \eref{eq:app_detrho} with the first square brackets omitted.
Inserting this back into \eqs (\ref{eq:app_Z2})-(\ref{eq:app_detA}) is shown in \cite{McCoy} to yield
\begin{equation}
-\frac{1}{\beta}\ln Z= 4 L (2M+1) F + L \mathcal{F}_0 +  L \mathcal{F}(h_\downarrow)+\mathcal{O}(1),
\end{equation}
where $F$ is given in \eref{eq:Fbulk_def},
\begin{equation}
\fl \mathcal{F}(h)= -\frac{1}{\beta} \Big\{ \ln [ \cosh(\beta h)] +\frac{1}{4\pi } \int_{-\pi}^\pi d\theta \ln \big[
1- \frac{z_h |1+e^{i\theta}|^2}{z_J(1+z^2_J+2z_J\cos \theta)-(1-z^2_J)\alpha} \big]\Big\}
\end{equation}
and $\alpha$ is given in \eref{eq:alpha_def}.
Since the two square brackets in \eref{eq:app_detrho} have the same structure, it is clear that the first square bracket would yield the same free energy, yielding
in the two-boundary-fields case the following free energy:
\begin{equation}
-\frac{1}{\beta}\ln Z= 4 L (2M+1) F + 2 L \mathcal{F}_0 +  L \mathcal{F}(h_\downarrow)+  L \mathcal{F}(h_\uparrow)+\mathcal{O}(1).
\end{equation}

\bibliographystyle{unsrt}
\bibliography{drivenising}

\end{document}